\documentclass[twocolumn,aps,showpacs]{revtex4}

\usepackage{epsfig}
\usepackage{lineno}
\usepackage{placeins}
\usepackage{subfigure}
\usepackage[ansinew]{inputenc}
\usepackage{multirow}
\usepackage{booktabs}
\usepackage{subscript}
\begin{document}

\title{Sub-threshold production of K$^0_s$ mesons and $\Lambda$ hyperons in Au(1.23A GeV)+Au}

\author{J.~Adamczewski-Musch$^{4}$, O.~Arnold$^{10,9}$, C.~Behnke$^{8}$, A.~Belounnas$^{16}$,
	A.~Belyaev$^{7}$, J.C.~Berger-Chen$^{10,9}$, J.~Biernat$^{3}$, A.~Blanco$^{2}$, C.~~Blume$^{8}$,
	M.~B\"{o}hmer$^{10}$
% ...
}

\affiliation{
	(HADES collaboration) \\
	Y.~Leifels$^{4}$\\
	\mbox{$^{1}$Institute of Physics, Slovak Academy of Sciences, 84228~Bratislava, Slovakia}\\
	\mbox{$^{2}$LIP-Laborat\'{o}rio de Instrumenta\c{c}\~{a}o e F\'{\i}sica Experimental de Part\'{\i}culas , 3004-516~Coimbra, Portugal}\\
	\mbox{$^{3}$Smoluchowski Institute of Physics, Jagiellonian University of Cracow, 30-059~Krak\'{o}w, Poland}\\
	\mbox{$^{4}$GSI Helmholtzzentrum f\"{u}r Schwerionenforschung GmbH, 64291~Darmstadt, Germany}\\
	\mbox{$^{5}$Technische Universit\"{a}t Darmstadt, 64289~Darmstadt, Germany}\\
	\mbox{$^{6}$Institut f\"{u}r Strahlenphysik, Helmholtz-Zentrum Dresden-Rossendorf, 01314~Dresden, Germany}\\
	\mbox{$^{7}$Joint Institute of Nuclear Research, 141980~Dubna, Russia}\\
	\mbox{$^{8}$Institut f\"{u}r Kernphysik, Goethe-Universit\"{a}t, 60438 ~Frankfurt, Germany}\\
	\mbox{$^{9}$Excellence Cluster 'Origin and Structure of the Universe' , 85748~Garching, Germany}\\
	\mbox{$^{10}$Physik Department E62, Technische Universit\"{a}t M\"{u}nchen, 85748~Garching, Germany}\\
	\mbox{$^{11}$II.Physikalisches Institut, Justus Liebig Universit\"{a}t Giessen, 35392~Giessen, Germany}\\
	\mbox{$^{12}$Institute for Nuclear Research, Russian Academy of Science, 117312~Moscow, Russia}\\
	\mbox{$^{13}$Institute of Theoretical and Experimental Physics, 117218~Moscow, Russia}\\
	\mbox{$^{14}$National Research Nuclear University MEPhI (Moscow Engineering Physics Institute), 115409~Moscow, Russia}\\
	\mbox{$^{15}$Department of Physics, University of Cyprus, 1678~Nicosia, Cyprus}\\
	\mbox{$^{16}$Institut de Physique Nucl\'{e}aire, CNRS-IN2P3, Univ. Paris-Sud, Universit\'{e} Paris-Saclay, F-91406~Orsay Cedex, France}\\
	\mbox{$^{17}$Nuclear Physics Institute, The Czech Academy of Sciences, 25068~Rez, Czech Republic}\\
	\mbox{$^{18}$LabCAF. F. F\'{\i}sica, Univ. de Santiago de Compostela, 15706~Santiago de Compostela, Spain}\\
	\\
	\mbox{$^{a}$ also at Coimbra Polytechnic - ISEC, ~Coimbra, Portugal}\\
	\mbox{$^{b}$ also at ExtreMe Matter Institute EMMI, 64291~Darmstadt, Germany}\\
	\mbox{$^{c}$ also at Technische Universit\"{a}t Dresden, 01062~Dresden, Germany}\\
	\mbox{$^{d}$ also at Frederick University, 1036~Nicosia, Cyprus}\\
	\mbox{$^{e}$ also at Dipartimento di Fisica and INFN, Universit\`{a} di Torino, 10125~Torino, Italy}\\
	\mbox{$^{\dag}$ deceased}\\
}

\date{18.12.2018}

\begin{abstract}
We present first data on sub-threshold production of K$^0_s$ mesons and $\Lambda$ hyperons in Au+Au collisions at $\sqrt{s_{\rm{NN}}}=2.4$ GeV. We observe an universal $\langle A_{part} \rangle$ scaling of hadrons containing strangeness, independent of their corresponding production thresholds. Comparing the yields, their $\langle A_{part} \rangle$ scaling, and the shapes of the rapidity and the $p_t$ spectra to state-of-the-art transport model (UrQMD, HSD, IQMD) predictions. We find that none of them can simultaneously describe these observables with reasonable $\chi^2$ values. 
\end{abstract}

\maketitle
\begin{figure}
	% Use the relevant command for your figure-insertion program
	% to insert the figure file.
	% For example, with the option graphics use
	\begin{center}
		\resizebox{8.cm}{!}{%
			
			\includegraphics{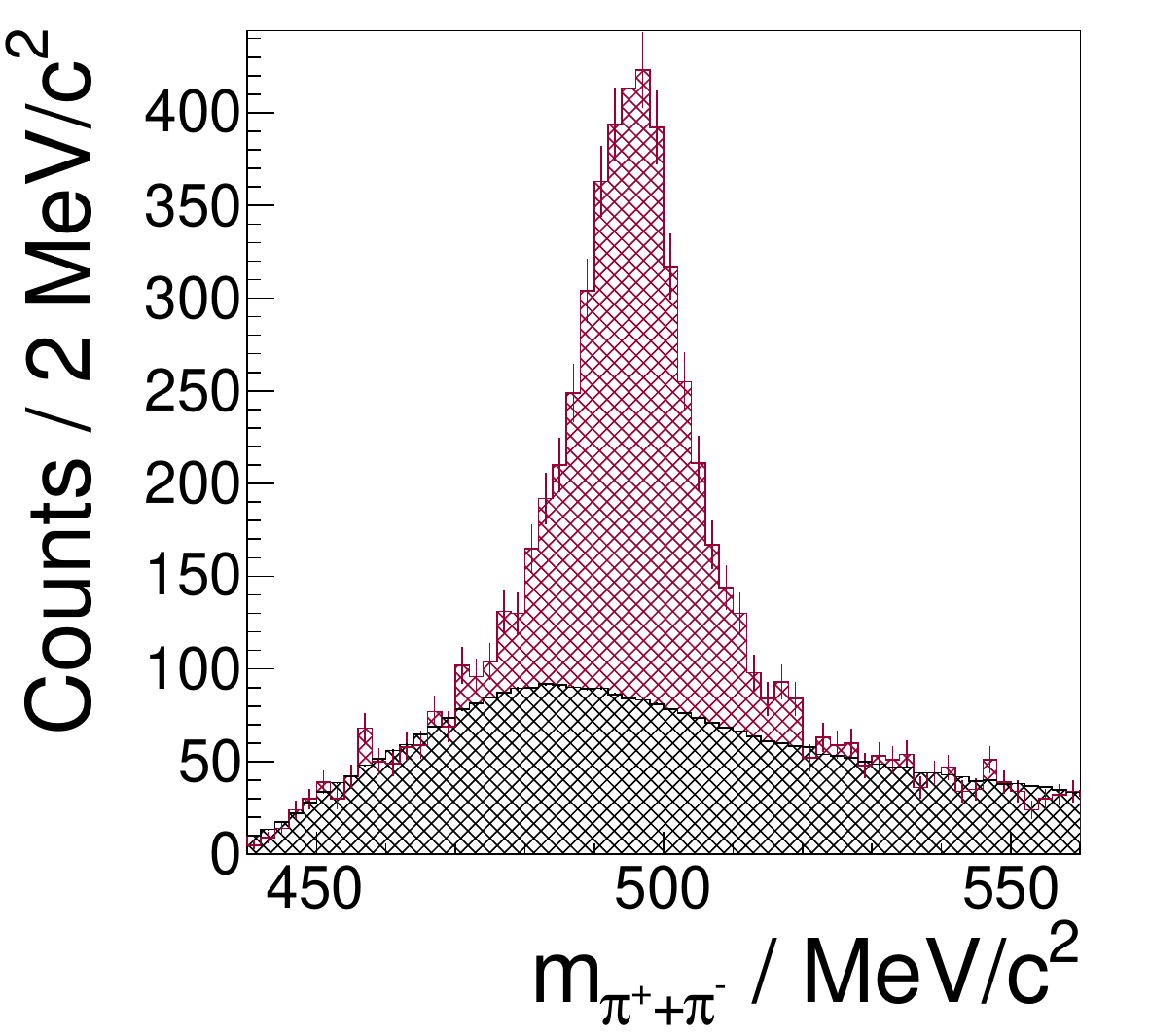}			
			\includegraphics{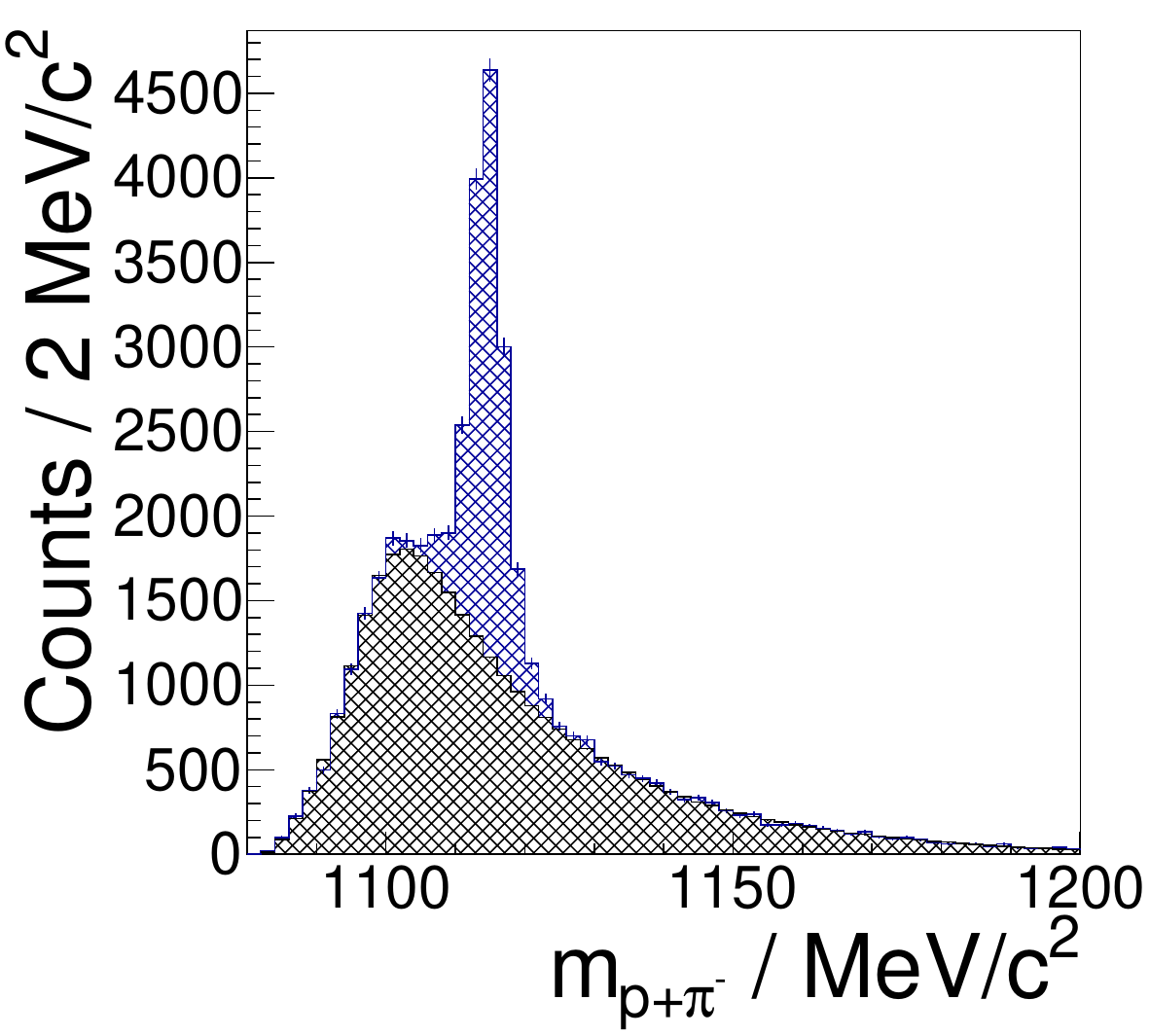}
			% epja impact_parameter_herb.eps!!
		}
	\end{center}
	% If not, use
	\vspace{-0.7cm}       % Give the correct figure height in cm
	\caption{Examples of K$^0_s$ (left) and $\Lambda$ (right) signals for 0-40\% most central events, over mixed-event background for the bin $-0.05<y_{cm}<0.05$ and reduced transverse masses between 80-120 MeV/$c^2$ and 100-150 MeV/$c^2$, respectively.}
	\label{signal}       % Give a unique label  
\end{figure}
\begin{figure}
		\begin{center}
			\resizebox{8.7cm}{!}{%				
				\includegraphics{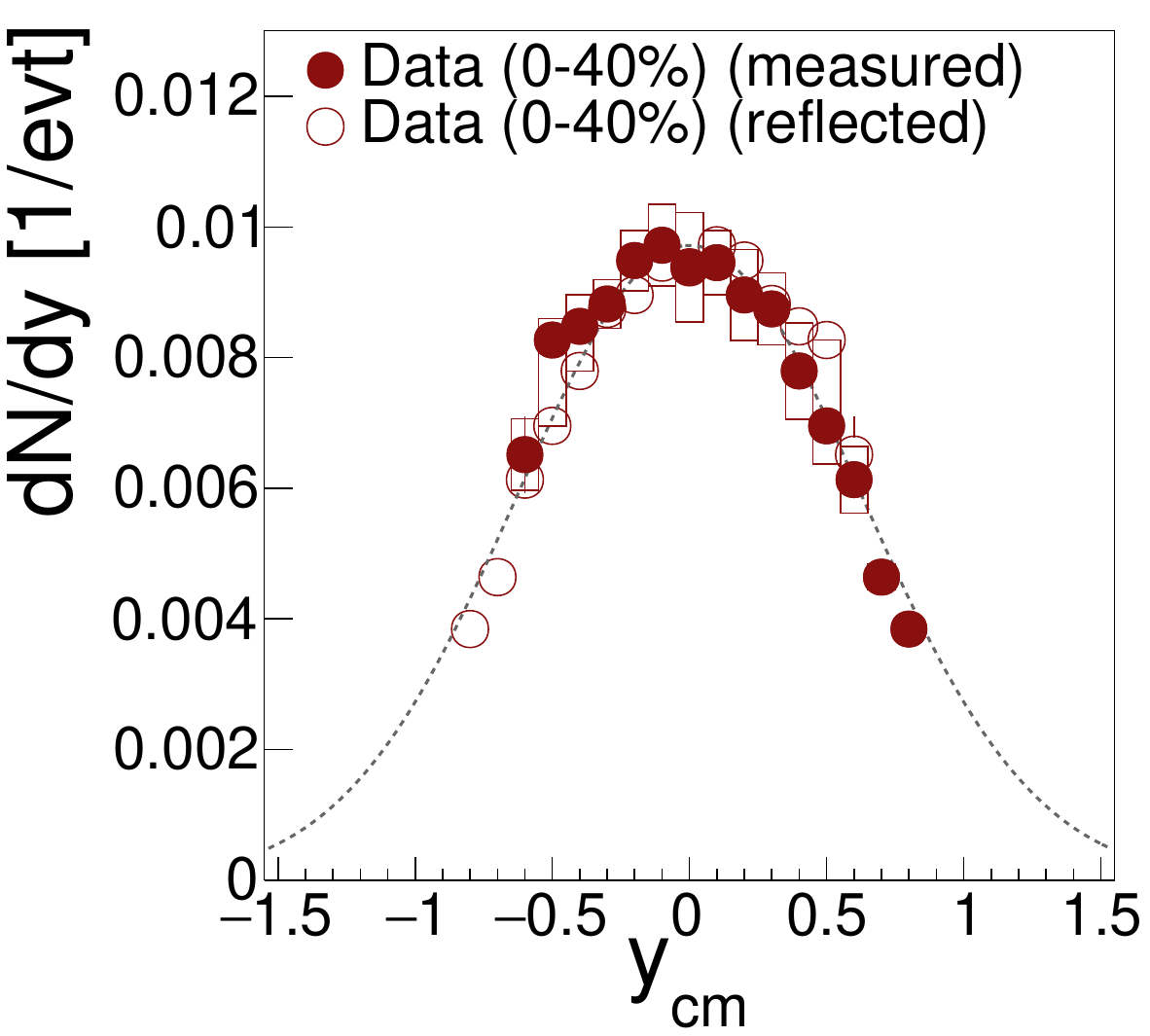}
				\includegraphics{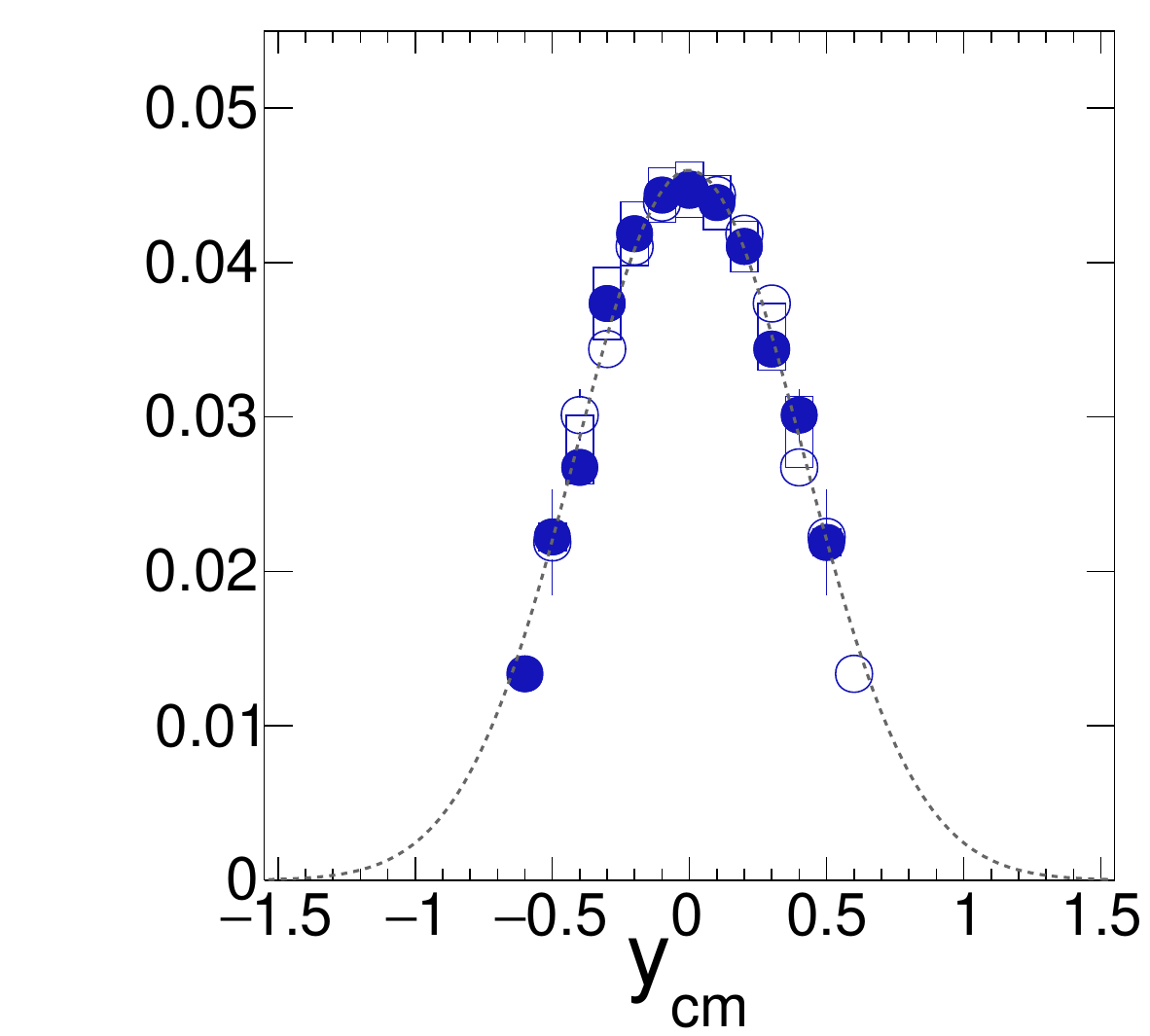}			
				% epja impact_parameter_herb.eps!!
			}
		\end{center}
		\vspace{-0.7cm}  
	\caption{Rapidity distribution of $K^0_s$ (left) and $\Lambda$ (right). The closed
		symbols depict the measured data points, whereas the open symbols show the data points
		reflected around the center-of-mass rapidity $y_{cm}=0$. The error bars display the statistical errors while the systematic uncertainties are indicated by the open boxes. For the extrapolation to
		unmeasured rapidity values a Gaussian function is used (dotted curve).}
	\label{fig_dNdy}       % Give a unique label
\end{figure}
\begin{figure}
	\begin{center}
		\resizebox{6.2cm}{!}{%				
			\includegraphics{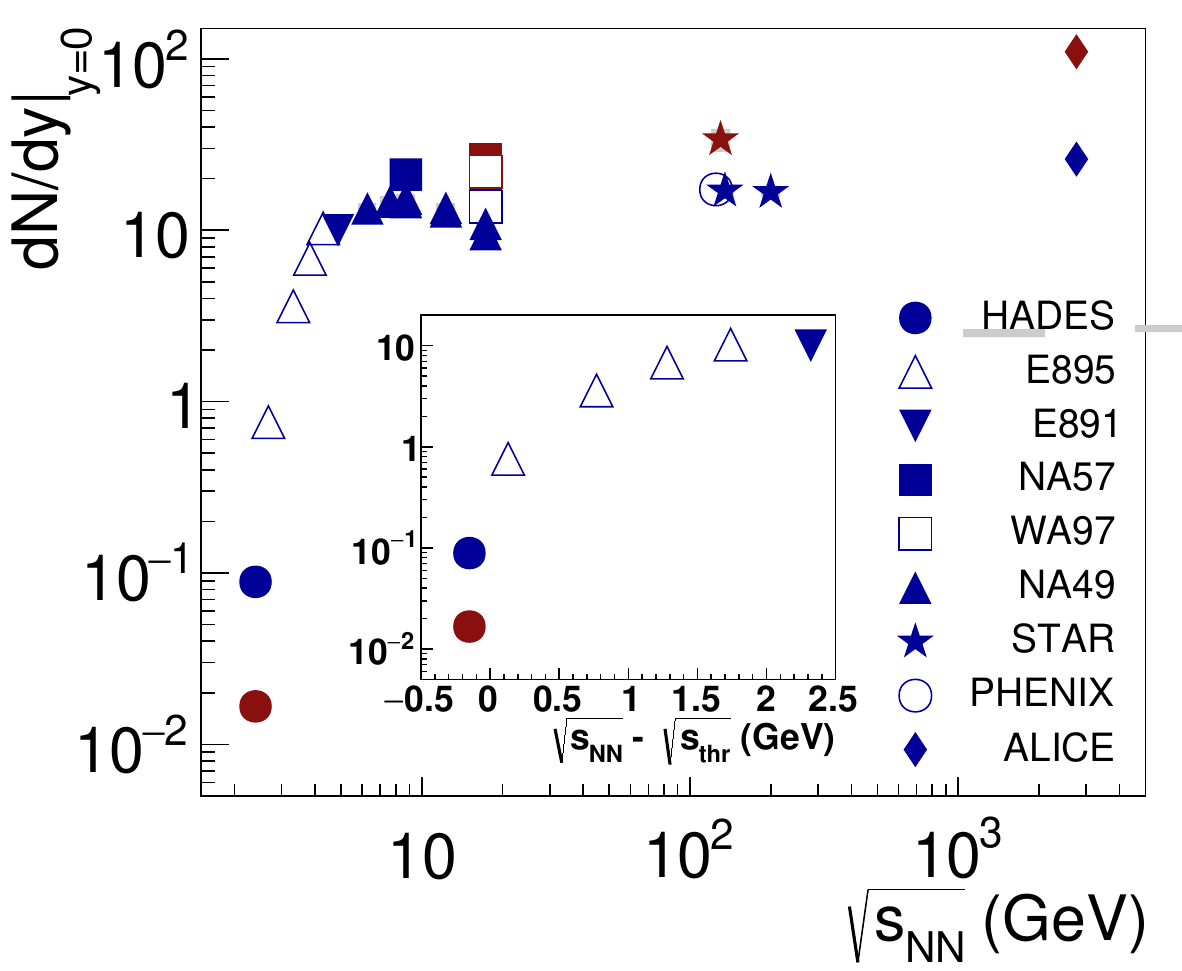}			
			%\includegraphics{plots/Lsqrt.pdf}
			% epja impact_parameter_herb.eps!!
		}
	\end{center}
	\vspace{-0.7cm}  
	\caption{Compilation of mid-rapidity yields for central Au+Au (Pb+Pb) collisions as a function of $\sqrt{s_{\rm{NN}}}$ of $K^0_s$ (red) and  $\Lambda$ (blue) from \cite{KNA57,KWA97,KSTAR,Blume:2011sb,Pinkenburg:2001fj,Alb,KWA97,LE891, LNA57,LPHENIX,LSTAR,LALICE}. Our results are shown by the two left most bullets.  %, of the form: $f(\sqrt{s})=C[1-(D/\sqrt{s})^G]^H$
    }
	\label{fig_ex}       % Give a unique label
\end{figure}

%\section{Introduction}
\renewcommand{\thefootnote}{\roman{footnote}}
\label{intro}
Relativistic heavy-ion collisions (HICs) provide a unique opportunity to study matter at 2-3 times nuclear ground state density (similar as expected for neutron star mergers \cite{Hanauske,dilept}) in the laboratory. In particular, kaons and $\Lambda$ hyperons are promising probes with relevance for various astrophysical processes \cite{Nelson,Pandharipande:1971up,Bethe,Glendenning:1991es,Balberg:1997yw,Lee94}. However, HICs are highly dynamical processes and therefore it is difficult to directly address fundamental aspects. Numerous works investigated kaon production in HICs in the few-GeV energy regime in the past. Comparisons of experimental data (spectra and flow anisotropies) to transport model calculations seem to confirm a repulsive K-N potential \cite{Benabderrahmane:2008qs,Agakishiev:2010zw,Agakishiev:2014moo,Forster07,Hartnack11,Mishra:2004te,Pal:2000yc,Metag:2017yuh}, which has been predicted by various effective approaches \cite{Kolomeitsev:1995xz,Schaffner97,Lutz94,Cassing97}.
Furthermore, constraints for the equation-of-state (EOS) of nuclear matter have been deduced from kaon production, under the assumption of energy accumulation in sequential nucleon-nucleon collisions, e.g. NN$\rightarrow\Delta$N \cite{Sturm:2000dm,Fuchs:2000kp,Hartnack:2005tr}.\\
Data on $\Lambda$ production from HICs at low energies are scarce. At SIS18 energies only data from small collision systems are available \cite{Bastid:2007jz,Agakishiev:2010rs}. While the $\Lambda$-nucleon potential is known to be attractive at ground state densities from hypernuclei formation \cite{Batty:1997zp}, its density dependence therefore still remains vague \cite{Weissenborn}.\\
In this paper, we report the first observation of K$^0_s$ and $\Lambda$ hyperons emitted from central Au+Au collisions at $\sqrt{s_{\rm{NN}}}=2.4$ GeV. Both kaons and $\Lambda$ hyperons are produced about 150 MeV below their free NN-threshold and hence are sensitive to the energy dissipation in the collision system. We compare the scaling of the multiplicities as function of the centrality of the collisions to the previously published data on charged kaons and $\phi$ mesons \cite{PhiKAuAu} and the spectra and rapidity distributions to predictions from three state-of-the-art microscopic transport models (UrQMD, HSD, IQMD) \cite{UrQMD,Cassing:1999es,Harti}. Based on this we discuss the validity of the previously drawn conclusions about the K-N potential and the energy dissipation during the collision in light of the new data.\\ 
%\section{Detector setup and analysis}
The data have been collected with HADES, located at the GSI Helmholtz Center for Heavy Ion Research in Darmstadt, Germany. HADES is a charged-particle detector consisting of a 6-coil toroidal magnet centered around the beam axis and six identical detection sections located between the coils covering almost the full azimuthal angle. Each sector is equipped with a Ring-Imaging Cherenkov (RICH) detector followed by Mini-Drift Chambers (MDCs), two in front of and two behind the magnetic field, as well as a scintillator hodoscope (TOF) and a Resistive Plate Chamber (RPC). At the end of the system a forward hodoscope used for event plane determination is located. The RICH detector is used mainly for electron/positron identification, the MDCs are the main tracking detectors, while the TOF and RPC are used for time-of-flight measurements in combination with a diamond start-detector located in front of the 15-fold segmented target. The trigger is based on the hit multiplicity in the TOF covering a polar angle range between $45^{\circ}$ and $85^{\circ}$. A detailed description of the HADES detector is given in \cite{Agakishiev:2009am}.\\
In total, $2.2 \times 10^{9}$ Au+Au events are used in the present analysis corresponding to the $40\%$ most central events. The latter is estimated based on studies using a Glauber model \cite{Kardan18}. \\
K$^0_s$ mesons are identified via their decay to $\pi^+$ and $\pi^-$ (BR =  69.2$\%$, c$\tau=2.68$ cm). $\Lambda$ hyperons are identified through their decay to $p$ and $\pi^-$ (BR =  63.9$\%$, c$\tau=7.89$ cm). Note that the reconstructed $\Lambda$ yield contains also a contribution from the (slightly heavier) $\Sigma^{0}$ baryon decaying electromagnetically exclusively into a $\Lambda$ and a photon. This decay process can not be detected with the present experimental setup; hence, the $\Lambda$ yield has to be understood as that of $\Lambda +\Sigma^0$ throughout the paper.
Pion and proton candidates used for the invariant mass analysis are identified by the curvature of their track in the magnetic field, a very loose cut on the reconstructed particle mass, and a careful track selection based on several quality parameters delivered by a Runge-Kutta tracking algorithm. Conditions required for the decay topology of both decay channels are used to suppress the combinatorial background of uncorrelated pairs. Cuts are applied on the distance between the primary event vertex and the decay vertex, on the distance of closest approach (DCA) between the proton, respectively the pion track, and the primary vertex, on the DCA between the two decay tracks, and on the DCA of the reconstructed mother particle trajectory to the primary vertex. Furthermore, a minimum opening angle is required \cite{timo}. For further suppression of the combinatorial background machine learning based on an artificial neural network \cite{tmva} is applied for recognition of weak decay topologies, in addition \cite{simon}. \\
The remaining background is subtracted using a mixed-event technique. Examples of the invariant mass distributions used for signal extractions are displayed in Fig. \ref{signal}. The signal counts are extracted by fitting a Gaussian and integrating the data in a $\pm2$ $\sigma$-region around the nominal mass, while the normalization region for the mixed-event background is placed between four-five $\sigma$ outside the signal region. Typical signal-to-background ratios are about 1-9 for K$^0_s$ and about 0.5-5 for $\Lambda$.   
In total, about 190000 K$^0_s$ mesons and about 290000 $\Lambda$ hyperons are reconstructed. \\
K$^0_s$ yields are determined in 15 rapidity bins, covering the center of mass rapidity $y_{cm}=y-0.74$ between -0.65 and +0.85 in steps of 0.1 units in rapidity, and up to 19 transverse mass ($m_{t}=\sqrt{p_{t}^{2}+m_{0}^{2}}$) bins in steps of 40 MeV/c$^{2}$. 
$\Lambda$ hyperons are identified in 12 rapidity bins, ranging from $y_{cm}=-0.65$ to +0.55  in steps of 0.1 units in rapidity, and up to 16 transverse mass bins in steps of 50 MeV/c$^{2}$.
\begin{figure}
	% Use the relevant command for your figure-insertion program
	% to insert the figure file.
	% For example, with the option graphics use
	\begin{center}
		\resizebox{8.7cm}{!}{%
			\includegraphics{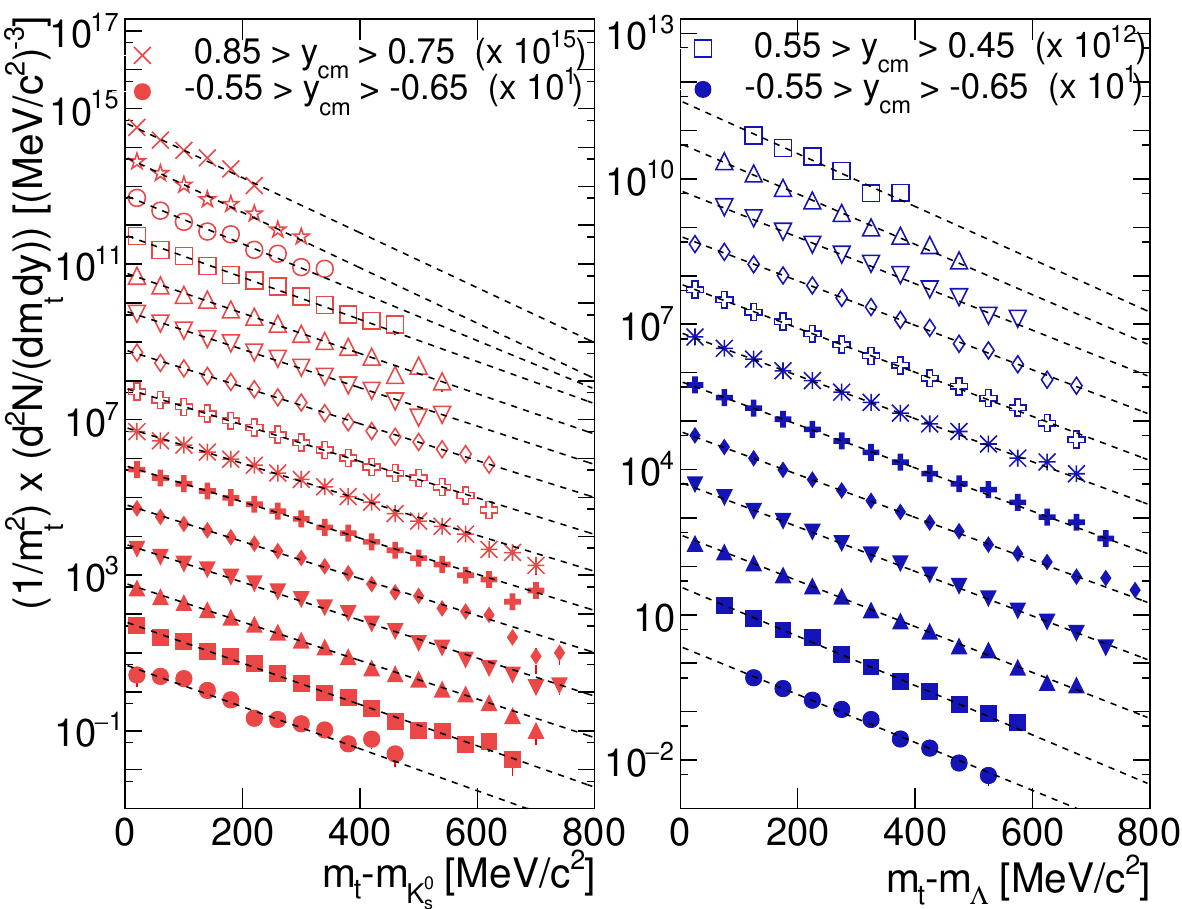}	
			% epja impact_parameter_herb.eps!!
		}
	\end{center}
	% If not, use
	\vspace{-0.7cm}       % Give the correct figure height in cm
	\caption{Reduced transverse mass ($m_{t}-m_{0}$) spectra of K$^0_s$ (left) and $\Lambda$ (right) for the 0-40\% most central events. For a better representation, the spectra are
		scaled by consecutive factors of 10 for each rapidity bin as indicated in the legend and only statistical errors are plotted. The dotted curves are fits with Eq.~\ref{bolz} to the data.}
	\label{mtK0}       % Give a unique label  
\end{figure}
The raw signal yields are corrected in each phase space cell for acceptance and efficiency using Monte-Carlo simulations based on Geant and a detailed description of the detector response, exposed to exactly the same reconstruction and analysis steps as the experimental data. As input for the simulation, thermal distributions of K$^{0}_{s}$ and $\Lambda$ hyperons with an inverse slope of 90 MeV were embedded into the experimental data.
The combined correction factors for the efficiency and acceptance correspond to about 50 for K$^{0}_{s}$ and about 100 for $\Lambda$ hyperons at mid-rapidity, including the branching ratio to the $\pi^+ \pi^-$ and $\pi^- p$ final state, respectively. In order to suppress the larger combinatorial background in the $p+\pi^-$ sample, more stringent cuts on the decay topology were applied than in case of the $\pi^+\pi^-$ sample, resulting in the lower detection efficiencies for the $\Lambda$ compared to the K$^{0}_{s}$ \cite{timo,simon}. \\  
%\section{Results}
The acceptance and efficiency corrected distributions of reduced transverse mass spectra for subsequent slices of rapidity for K$^{0}_{s}$ and $\Lambda$ hyperons are presented in Fig.~\ref{mtK0}. Displayed is the number of counts per event, per transverse mass and per unit in rapidity, divided by $m_t^2$. This representation is chosen to ease a comparison with single slope Boltzmann fits to the resulting distribution according to
\begin{equation}
\label{bolz}
\frac{1}{m_{t}^{2}} \frac{d^2N}{dm_{t}dy_{cm}} = C(y_{cm}) \,
\exp \left( -\frac{(m_t-m_0)c^2}{T_B(y_{cm})}  \right), 
\end{equation} 
which describe the spectra satisfactorily. 
The rapidity distributions, shown in Fig.~\ref{fig_dNdy} are obtained by integrating the data as function of the transverse momentum $p_{t}$ and using Boltzmann fit functions for extrapolations in the not covered $p_{t}$ regions.
The systematic errors of the yields in each rapidity bin are due to the variations of topology cuts, the normalization region of the mixed-event background and by the comparison of the spectra measured in the forward and backward hemisphere.
The decay length distributions of the two hadrons are determined to ensure the quality of the correction procedure. We observe lifetimes in agreement with the PDG values, i.e. K$^{0}_{s}$: $\tau_{exp}=87.1\pm1.1$~ps,  $\tau_{PDG}=89.6\pm1.6$ ps; $\Lambda$: $\tau_{exp}=255\pm7$ ps, $\tau_{PDG}=263\pm2$ ps \cite{PDG}.\\
Multiplicities are obtained by integrating the rapidity distribution and using a Gaussian fit for extrapolation to full phase space, see dotted curve in Fig.~\ref{fig_dNdy}. The statistical error is taken from the fit directly. The systematic uncertainty of the extrapolation is estimated based on variation within the systematic errors and using in addition to the Gaussian fit the rapidity distributions obtained from the three different transport models for extrapolation, as described below.\\
We obtain a total multiplicity of $(1.56 \pm 0.03_{stat} {}^{+ 0.12}_{- 0.12}{}_{sys}) \times 10^{-2}$ K$^0_s$ and $(4.72 \pm 0.06_{stat} {}^{+ 0.21}_{- 0.75}{}_{sys}) \times 10^{-2}$ $\Lambda$.\\
%The rapidity distributions and the Gaussian functions used for extrapolation in rapidity are displayed in Fig. \ref{fig_dNdy}. The error bars display the statistical error while the systematic uncertainty is indicated by the open boxes.\\
The extracted inverse slope parameters obtained from the Boltzmann fits to the $m_{t}$ spectra for each rapidity interval are fitted using the ansatz $T_{B}=\frac{T_{eff}}{\cosh(y_{cm})}$ in order to obtain the effective inverse slope $T_{eff}$. As the inverse slope contains a contribution from the velocity of the radial expansion of the fireball, which is proportional to the particle mass, one expects a larger inverse slope for the $\Lambda$ hyperons. We find $T_{eff}=93\pm1\pm4$ MeV for the K$^{0}_{s}$ and $T_{eff}=98\pm1\pm4$ for the $\Lambda$ hyperons, suggesting no strong difference between radial flow of the K$^{0}_{s}$ and $\Lambda$ hyperons.\\
\begin{table}
	\begin{center}
		\begin{tabular}{|c|c|c|c|} \hline
			\textbf{K$^0_s$} & yield $\times$ 10$^{2}$ [1/evt] & $T_{eff}$ [MeV] \\ \hline \hline
			0 - 40\% & 1.56 $\pm$ 0.03 \textsuperscript{+ 0.12}\llap{\textsubscript{$-$\hspace{-0.14px} 0.12}}  & 93 $\pm$ 1 $\pm$ 4\\ \hline
			0 - 10\% & 2.84 $\pm$ 0.09 \textsuperscript{+ 0.21}\llap{\textsubscript{$-$\hspace{-0.14px} 0.27}}  & 98 $\pm$ 1 $\pm$ 3\\ %\hline
			10 - 20\% & 1.58 $\pm$ 0.04 \textsuperscript{+ 0.12}\llap{\textsubscript{$-$\hspace{-0.14px} 0.12}} & 93 $\pm$ 1 $\pm$ 3\\ %\hline
			20 - 30\% & 1.06 $\pm$ 0.03 \textsuperscript{+ 0.08}\llap{\textsubscript{$-$\hspace{-0.14px} 0.08}} & 89 $\pm$ 1 $\pm$ 1\\ %\hline
			30 - 40\% & 0.66 $\pm$ 0.02 \textsuperscript{+ 0.06}\llap{\textsubscript{$-$\hspace{-0.14px} 0.06}} & 86 $\pm$ 1 $\pm$ 1\\ \hline  \hline 
			\textbf{$\Lambda$} & yield $\times$ 10$^{2}$ [1/evt] & $T_{eff}$ [MeV] \\ \hline \hline
			0 - 40\% & 4.72 $\pm$ 0.06 \textsuperscript{+ 0.21}\llap{\textsubscript{$-$\hspace{-0.14px} 0.75}} & 98 $\pm$ 1 $\pm$ 5\\ \hline
			0 - 10\% & 8.22 $\pm$ 0.11 \textsuperscript{+ 0.55}\llap{\textsubscript{$-$\hspace{-0.14px} 0.92}} & 106 $\pm$ 1 $\pm$ 2\\ %\hline
			10 - 20\% & 4.90 $\pm$ 0.09 \textsuperscript{+ 0.21}\llap{\textsubscript{$-$\hspace{-0.14px} 0.7}} & 97 $\pm$ 1 $\pm$ 2\\ %\hline
			20 - 30\% & 3.17 $\pm$ 0.08 \textsuperscript{+ 0.14}\llap{\textsubscript{$-$\hspace{-0.14px} 0.36}}& 90 $\pm$ 1 $\pm$ 3\\ %\hline
			30 - 40\% & 1.92 $\pm$ 0.08 \textsuperscript{+ 0.09}\llap{\textsubscript{$-$\hspace{-0.14px} 0.28}}& 84 $\pm$ 1 $\pm$ 4\\ \hline   
		\end{tabular}
	\end{center}
	\vspace{-0.5cm}
	\caption{K$^0_s$ and $\Lambda$ multiplicities in full phase space and inverse slopes at mid-rapidity $T_{eff}$ for a given centrality. The first given error corresponds always to the statistical, the second to the systematic error. See text for details.}
	\label{tab}
\end{table}
In addition, the analysis procedure is repeated in the same way for four centrality classes. These classes correspond to 10$\%$ steps in centrality, which can be translated into the average number of participants $\langle A_{part} \rangle$ \cite{Kardan18}. %which translate to $\langle A_{\rm{part}} \rangle$ of 301$\pm$11, 212$\pm$10, 149$\pm$8, 102$\pm$6. 
The results are summarized in Tab. \ref{tab}.\\ 
A comparison to the world data is presented in Fig.~\ref{fig_ex}, where the mid-rapidity yields for central Au+Au (Pb+Pb) collisions as function of $\sqrt{s_{\rm{NN}}}$ are displayed. While only experimental data on K$^0_s$ production exists for central HICs at energies $\sqrt{s_{\rm{NN}}}>17.2$ GeV \cite{KNA57,KWA97,KSTAR,Blume:2011sb}, the $\Lambda$ hyperon was studied more extensively \cite{Pinkenburg:2001fj,Alb,KWA97,LE891, LNA57,LPHENIX,LSTAR,LALICE}. Its yield rises almost exponentially with energy up to $\sqrt{s_{\rm{NN}}}\approx$ 5 GeV and then levels off. \\
%For $\Lambda$ we find that the measured yield at mid-rapidity fits well to systematics extrapolated from higher energies, for the K$^0_s$ it constrains the low-$\sqrt{s}$ behavior.
%The observed behavior can be described in both cases with a simple parameterization up to top RHIC energy.  %$f(\sqrt{s})=C[1-(D/\sqrt{s})^G]^H$. 
%In case of the $\Lambda$, the world data at higher energies constrain the fit sufficiently to extrapolate down to SIS18 energies. We find our point fitting well into this extrapolation. In case of the $K^0_s$ however, our point is necessary to constrain the low energy behavior. In both cases our simple parameterization does not cover the rise for energies larger 10 GeV and hence misses the data points at LHC energies. \\
\begin{figure}[b]
	% Use the relevant command for your figure-insertion program
	% to insert the figure file.
	% For example, with the option graphics use
	\begin{center}
		\resizebox{5.2cm}{!}{%
			\includegraphics{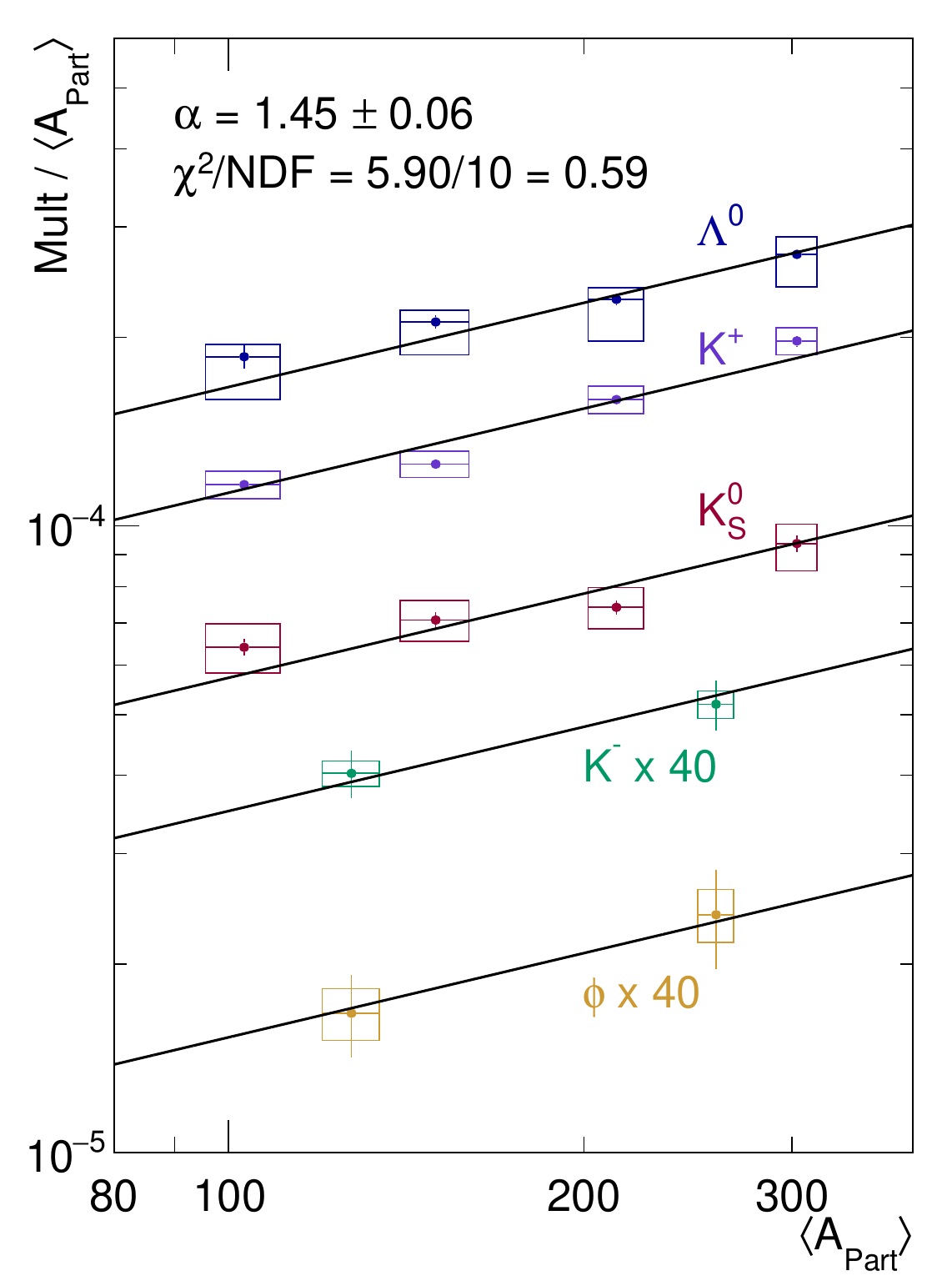}
			%\includegraphics{plots/KKroland.pdf}
			%\includegraphics{plots/phiKroland.pdf}
			% epja impact_parameter_herb.eps!!
		}
	\end{center}		
	% If not, use
	\vspace{-0.7cm}       % Give the correct figure height in cm
	\caption{Multiplicities per mean number of participants Mult/$\langle A_{part} \rangle$ as a function of $\langle A_{part} \rangle$. All hadron yields are fitted simultaneously with a function of the form Mult $\propto \langle {A_{part}} \rangle ^\alpha$ with the result: $\alpha=1.45\pm0.06$.}
	\label{Apart}       % Give a unique label
\end{figure}
As both the $K^0_s$ and $\Lambda$ are produced below their free NN-threshold, the required energy must be supplied from the collision system. Hence, one expects their yields to rise as a function of the geometrical overlap of the nuclei, which is an approximate of the number of nucleons taking part in the collision.
To investigate this, we analyse the multiplicities per mean number of participants Mult/$\langle A_{part} \rangle$ as a function of $\langle A_{part} \rangle$, as shown in Fig.~\ref{Apart} and include also the multiplicities of charged kaons and $\phi$ mesons measured in the same collision system \cite{PhiKAuAu}. If final state interactions play a minor role, the strength of this rise characterizes the amount of surplus energy provided by the system, over the contribution from first chance NN collisions. 
If one assumes that energy accumulates in sequential nucleon-nucleon collisions (as discussed in the introduction) in combination with the steep energy excitation function for strange hadron production, one expects to observe significantly different slopes, due to the clear hierarchy in the production thresholds, $\approx$ -150 MeV for K$^+$, K$^0$, $\Lambda$ (NN$\rightarrow$N$\Lambda$K) and $\approx$ -450 MeV, $\approx$ -490 MeV for the K$^-$ (NN$\rightarrow$NN$K^+K^-$) and the $\phi$ meson (NN$\rightarrow$NN$\phi$). However, the global fit of the function Mult $\propto \langle {A_{part}} \rangle ^\alpha$ to all the hadron yields returns a satisfactory value of $\chi^2$/NDF = 0.59, with $\alpha$ = 1.45 $\pm$ 0.06 \footnotemark[1]\footnotetext[1]{While in case of the $K^-$ the similar scaling was explained by a coupling to the $K^{+,0}$ yield via strangeness exchange reactions, e.g. $\pi^0+\Lambda\rightarrow K^-+p$ \cite{Forster07}, no such process is possible in case of the $\phi$ meson.}. This points to a more involved picture than assumed in the past, as the total amount of produced strangeness increases with the number of participants and might be only redistributed (statistically) to the final hadron species at freeze-out \cite{kolo}. This implies that the created system is more interrelated than expected in the past.\\
\begin{figure}
	\begin{center}
		\resizebox{8.7cm}{!}{%				
			\includegraphics{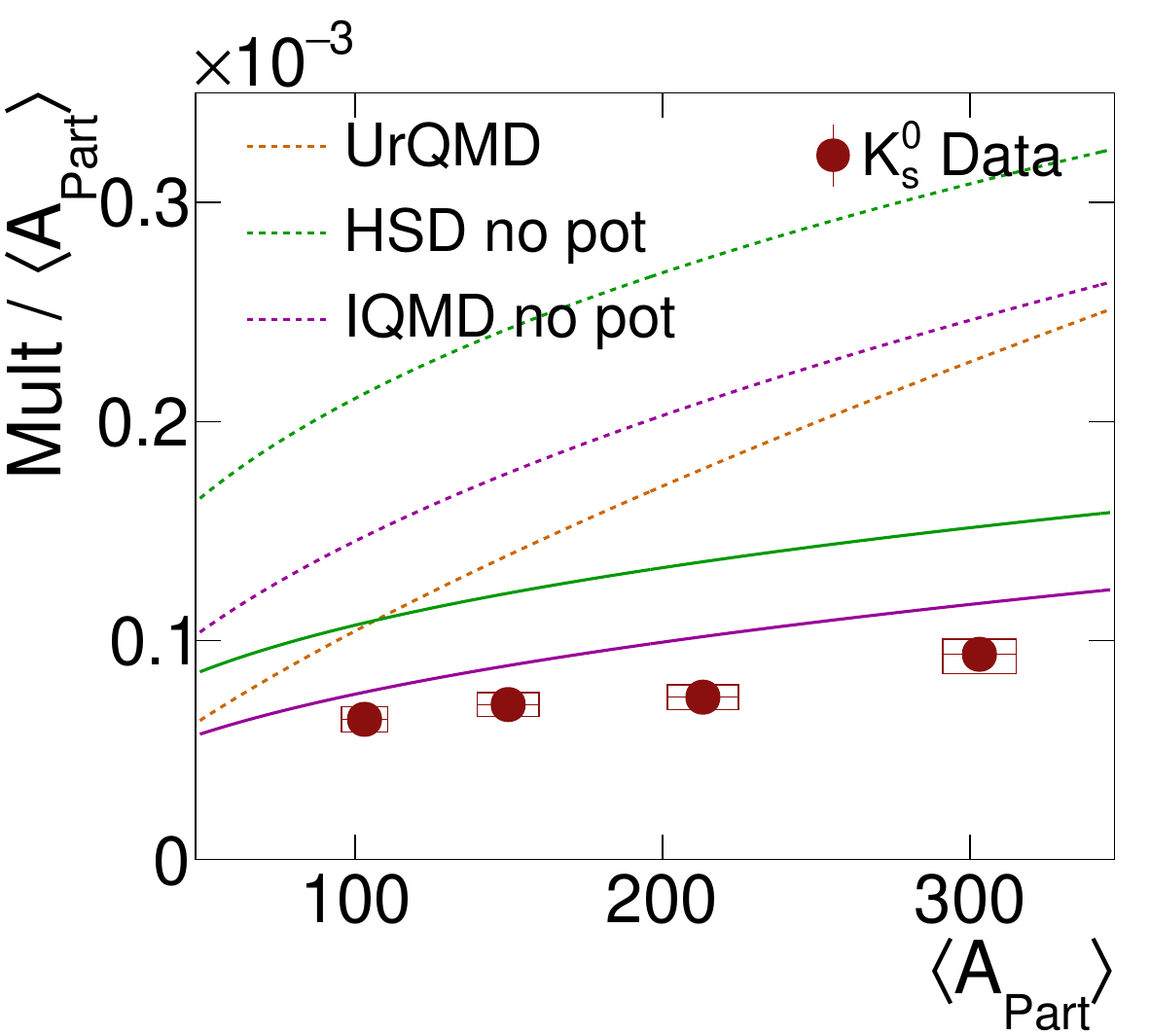}
		    \includegraphics{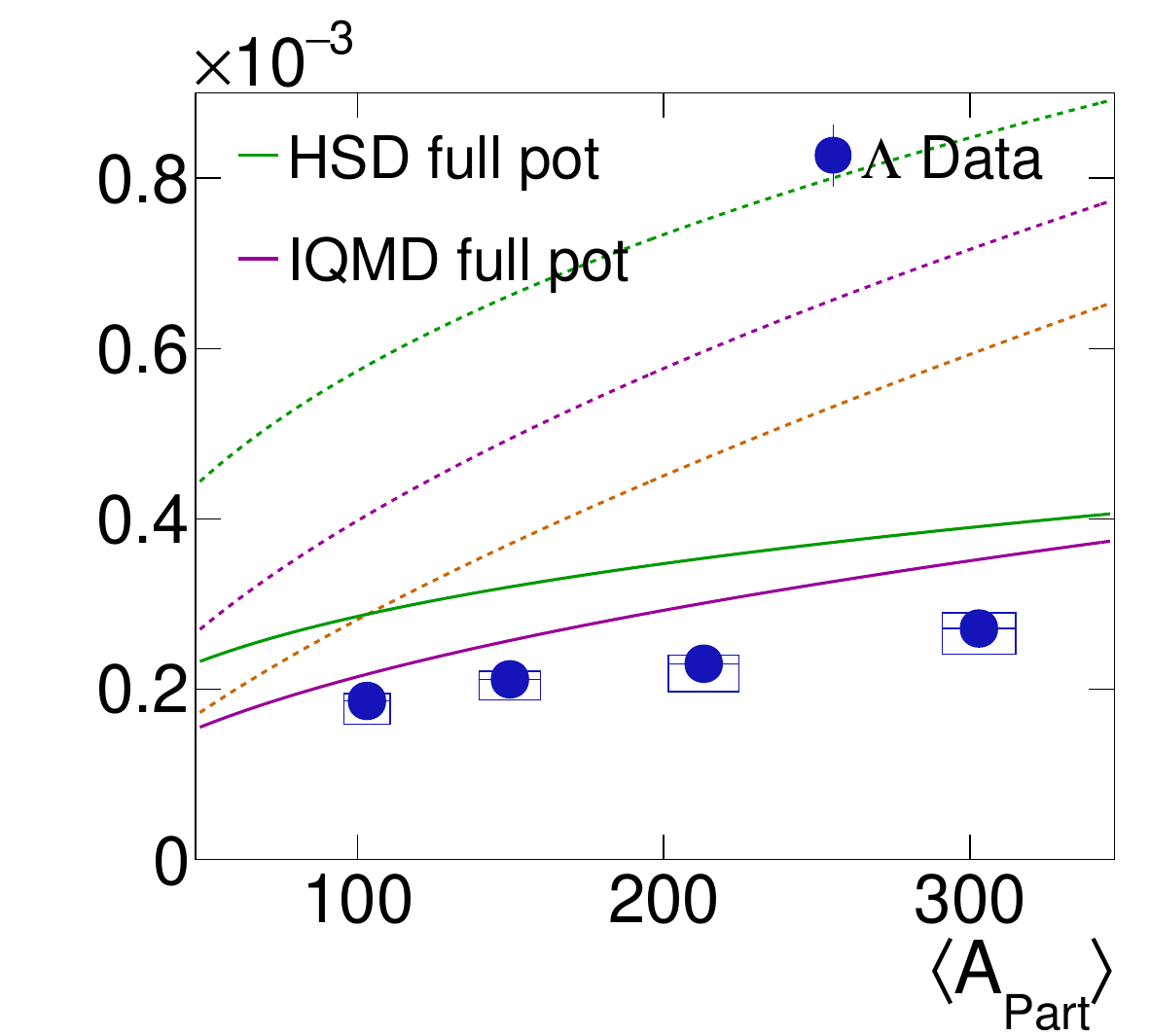}				
			% epja impact_parameter_herb.eps!!
		}
	\end{center}
	\vspace{-0.7cm}  
	\caption{Multiplicities per mean number of participants Mult/ $\langle A_{part} \rangle$, as a function of $\langle A_{part} \rangle$ for K$^0_s$ (left) and  $\Lambda$ (right) compared to various transport model calculations (see legends).}
	\label{fig_alpha}       % Give a unique label
\end{figure} 
\begin{table}
	\begin{center}
		\begin{tabular}{|c|c|} \hline
			&$\alpha$ \\ \hline 
			data ($K^{+,-,0}$, $\Lambda$, $\phi$) & 1.45$\pm$0.06   \\ \hline
			UrQMD ($K^0_s$, $\Lambda$) & 1.69$\pm$0.04 \\ \hline
			HSD no pot. ($K^0_s$, $\Lambda$)& 1.35$\pm$0.02 \\ \hline
			IQMD no pot. ($K^0_s$, $\Lambda$)& 1.51$\pm$0.03 \\ \hline
			HSD pot. ($K^0_s$, $\Lambda$)& 1.30$\pm$0.02 \\ \hline
			IQMD pot.($K^0_s$, $\Lambda$)& 1.42$\pm$0.03 \\ \hline		
		\end{tabular}
	\end{center}
	\vspace{-0.5cm}
	\caption{Values of the parameter $\alpha$ extracted for K$^0_s$ and $\Lambda$ data and various models, as displayed in Fig. \ref{fig_alpha}.}
	\label{tab2}
\end{table}
%\section{Model comparison}
\begin{figure}
	\begin{center}
		\resizebox{8.7cm}{!}{%				
			\includegraphics{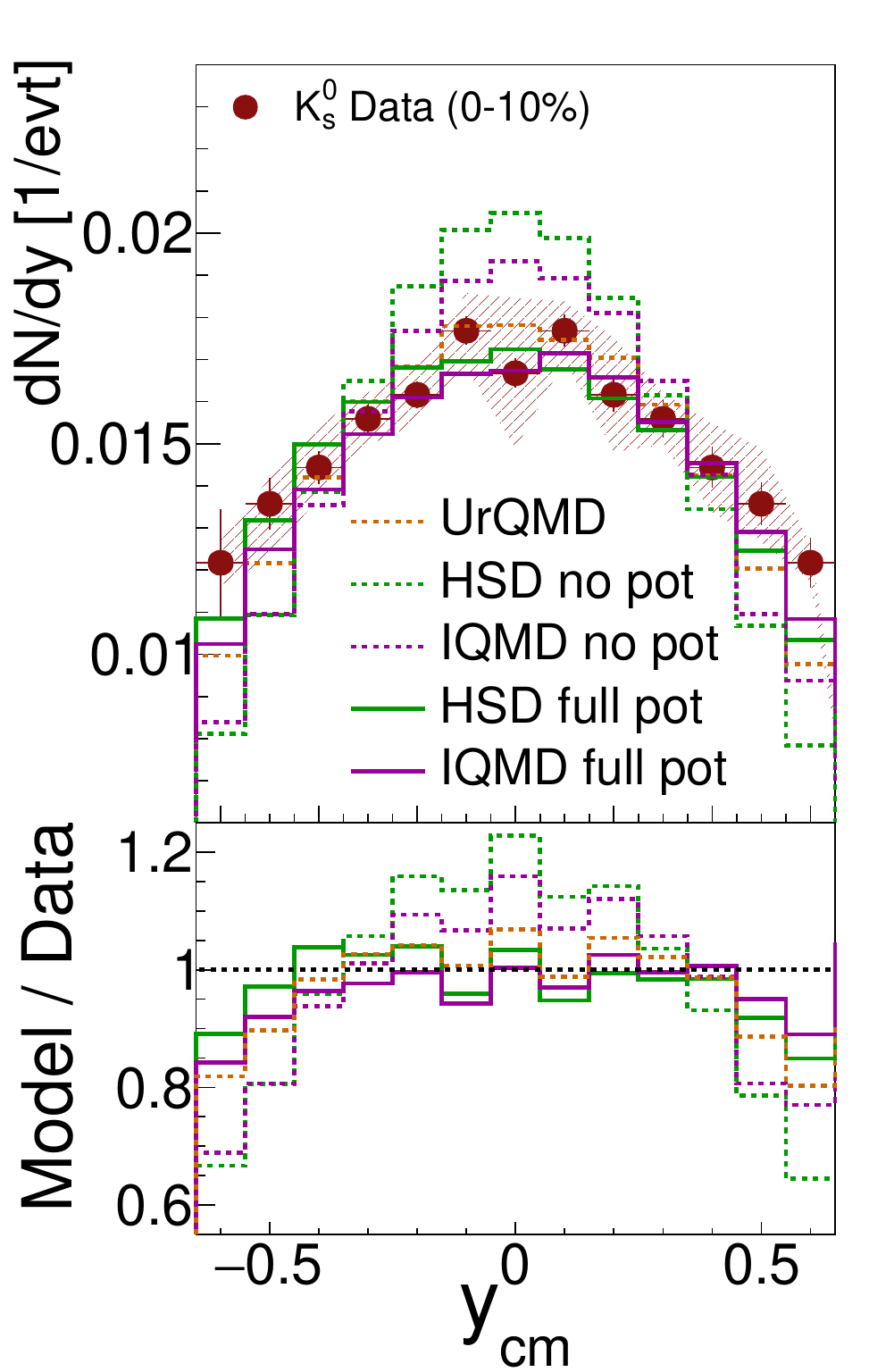}			
		    \includegraphics{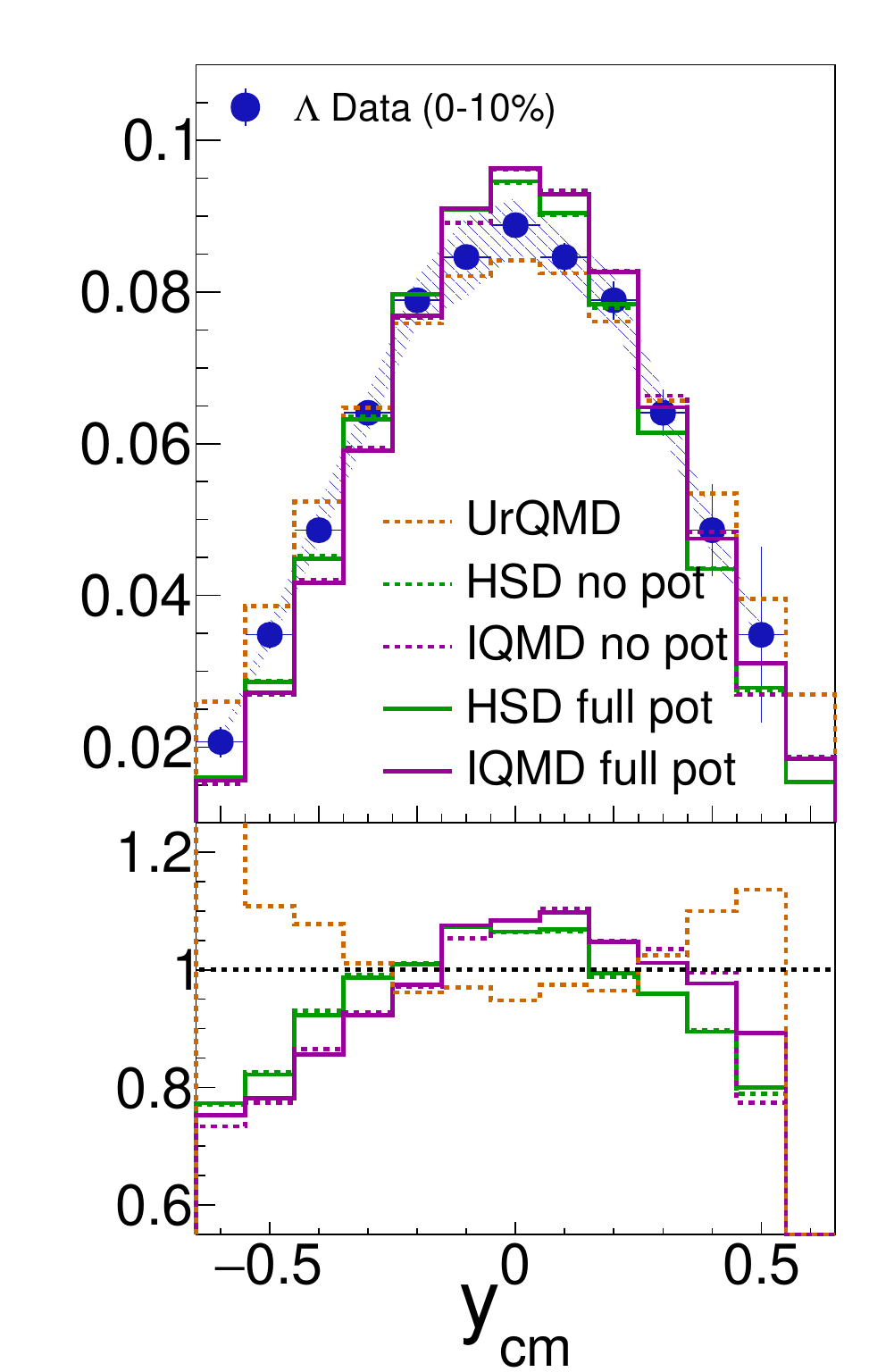}		
			% epja impact_parameter_herb.eps!!
		}
	\end{center}
	\vspace{-0.7cm}  
	\caption{Comparison of the shape of the rapidity distribution of K$^0_s$ (left) and  $\Lambda$ (right) to various transport model versions. The model curves are normalized to the integral of the data, see text for details. Data are symmetrized around mid-rapidity.}
	\label{fig_y}       % Give a unique label
\end{figure} 
In the following, we will compare the K$^0_s$ and $\Lambda$ data to predictions from three state-of-the-art hadronic transport models, the Isospin Quantum Molecular Dynamics model (IQMDv.c8) \cite{Harti}, the Hadron String Dynamics (HSDv.711n) model \cite{Cassing:1999es} and the Ultrarelativistic Quantum Molecular Dynamics model (UrQMDv 3.4) \cite{UrQMD}. It has been shown that none of the standard code versions can reproduce the observed $\phi$/$K^-$ multiplicity ratio measured in the same experiment \cite{Steinheimer:2015sha} \footnotemark[2]\footnotetext[2]{The UrQMD version predicting the measured $\phi$/$K^-$ \cite{PhiKAuAu} ratio after tuning the cross-sections to match data from p+p collisions is not publicly available yet \cite{Steinheimer:2015sha}.}.
All three are semi-classical models simulating a HIC on an event-by-event basis. While UrQMD produces particles via intermediate resonance excitations, in HSD and IQMD also direct production via two-to-three particle processes is included however in IQMD only $\Delta(1232)$ resonances are implemented. In contrast to IQMD and HSD, neither mean-field N-N potentials nor explicit K-N potentials are included in the presented version of UrQMD. IQMD is, due to missing high-energy processes, not applicable at energies beyond 2 $A$ GeV, but it is very well tested in the SIS18 energy regime. HSD allows in addition, the propagation of off-shell particles, however, this is more relevant for antikaon production. \\
%The created system in a HIC can be characterized by the time evolution of two macroscopic observables, the baryon density $\rho_B$ and the average kinetic energy per particle $\langle E_{kin} \rangle$. The latter one can be associated with a temperature, if particles experience a sufficient number of collisions.
%Both of these observables depend on various microscopic effects, which can compensate each other and are hence difficult to constrain and to differentiate, leading to possible ambiguities in the effective description of observables (for further reading, we refer e.g. to \cite{Andronic:2004}). The maximum density reached depends on the amount of stopping of participating nucleons in the reaction zone, as well as the repulsive two- and many-body forces between nucleons, which defines the compressibility of nuclear matter. The average kinetic energy per particle $\langle E_{kin} \rangle$ depends on the radial expansion velocity of the system and the amount of particles which can leave the collision zone without scattering, hence the cross-section between the particles. \\
We try to extract particle specific properties of K$^0_s$ mesons and $\Lambda$ hyperons  like the K-N and $\Lambda$-N potential, which affect both their production and propagation in the medium.\\ 
%All three models have been used to reproduce the bulk properties the created system of the created system sufficiently well in order to investigate the microscopic properties of K$^0_s$ and $\Lambda$.\\   
We start with the comparison of centrality dependence of the integrated yield, see Fig. \ref{fig_alpha} and Tab. \ref{tab2}. 
We find that HSD and IQMD without an implementation of the K-N potential, as well as UrQMD, overpredict the yields by a large factor. Also the model curves differ among each other by up to a factor 2.5, which points to the use of different parametrizations for elementary cross sections.\\ %In case of UrQMD the rise with $\langle A_{part} \rangle$ is over-predicted. % which is the only microscopic model, which predicted correctly the recently measured $\phi/K^-$ ratio in the same collision system after tuning the cross-sections to match data from p+p collisions \cite{Steinheimer:2015sha,PhiKAuAu}. 
%which might be due to wrongly estimated cross sections for the not well known intermediate high-mass resonances and their branching ratios to final states involving strangeness. These resonances are an effective way to redistribute energy in the system.
%The differences among the models amount to roughly 30\%.\\% which we consider as the minimal systematic uncertainty when comparing experimental yields and model predictions.\\     
In HSD and IQMD, a repulsive K-N potential of 40 MeV at nuclear ground state density $\rho_0$ is included, which increases linearly with density. If turned on, the K$^0_s$ curves come much closer to the data and also the $\alpha$ parameter is reduced. The IQMD predictions are by far the closest to the data with a deviation of the yields of the order of 10\% \footnotemark[3]\footnotetext[3]{Note, that preliminary data on yields of charged pions show a deviation at the order of 20\% to data of the FOPI experiment, which are well described by IQMD \cite{Reisdorf}. Due to $\pi$ induced strangeness production channels, this difference is also transported to the K$^0_s$ and $\Lambda$ yields.} and an agreement within errors of the extracted values of $\alpha$. The reduction of the yield and $\alpha$ values can be understood qualitatively by an effective shift in the production threshold of kaons. As the effect of the density dependence is more pronounced for central events, also the rise with $\langle {A_{part}} \rangle$ is reduced. Due to the associated production of kaons and $\Lambda$ hyperons, also the $\Lambda$ yields are affected by the inclusion of a K-N potential. Note that the employed version of UrQMD does not include any potential, while both versions of HSD and IQMD assume the strength of the $\Lambda$-N mean field to be 2/3 of the N-N mean field, motivated by the additive quark model \cite{Hartnack11}.\\
Next, we compare the shape of the rapidity distributions for 0-10\% most central events. For this, we have symmetrized the distributions with respect to mid-rapidity. In order to compare the shapes, the model curves are normalized to the area of the experimental ones, see Fig. \ref{fig_y}. The width of the rapidity distribution is particularly sensitive to the stopping of baryons in the collision zone. %One can pin this down to the question, how many times a nucleon must scatter, in order to gain enough energy for the production of a $K^0_s$-$\Lambda$ pair.
Repulsive potentials influence the shape of the distribution further by pushing the particles away from the bulk of matter at mid-rapidity. Indeed, we find that the inclusion of a potential improves the description significantly. In contrast to the previous observable, also the UrQMD calculation gives a fair description of  the shape, without any potential. In case of the $\Lambda$, the inclusion of the K-N potential does not affect the shape and we find that UrQMD describes the data best.\\  
Finally, we study the transverse momentum distribution at mid-rapidity for the most central event class, see Fig. \ref{fig_y}. Besides the production mechanism \cite{Steini} and the radial expansion velocity of the system, the low transverse momentum part is particularly sensitive to the K/$\Lambda$-N potentials \cite{Agakishiev:2010zw} \footnotemark[4]\footnotetext[4]{Note that this depends on the form of the implemented potential and hence does not hold true for all models, see e.g. \cite{Agakishiev:2014moo}.}. Once again, in order to compare the shapes, the model curves are normalized to the area of the experimental ones. Clearly, for K$^0_s$ the data  favour models which include potentials. However, again UrQMD without any potential shows a completely different behavior compared to the two other calculations without the potential, undershooting the low $p_{t}$ part of the spectrum. Hence, production via intermediate resonances seems to (over-) mimic the effect of the potential.\\
%Therefore, we normalize the spectra in the high transverse momentum part ($600 < p_t < 900 $ MeV/c for K$^0_s$ and $700 < p_t < 1000 $ MeV/c for $\Lambda$, respectively). Clearly, for K$^0_s$ the data  favour models which include potentials; the IQMD prediction describes the shape very well. However, again UrQMD without any potential shows a completely different behavior at low $p_{t}$, compared to the two other calculations without the potential. Hence, production via intermediate resonances seems to (over-) mimic the effect of the potential.\\
In addition, similar as in case of the rapidity distribution, UrQMD offers the best description of the $\Lambda$ $p_t$ spectra.\\ 
In total, we find that none of the model predictions can  describe the yield, the $\langle A_{part} \rangle$ scaling and the shape of the rapidity and $p_t$ spectra of K$^0_s$ and $\Lambda$ simultaneously, see also the $\chi^2$ values normalized to the number of data points listed in Tab. \ref{tabsum} for the investigated observables. Furthermore, we observe that effects of a repulsive K-N potential can be to some extend compensated by production via intermediate resonances.
Hence, any further conclusions are weakened by the ambiguities of different microscopic effects and the incomplete description of the presented observables within all investigated models. Therefore, it is misleading to draw conclusions on the strength of the potentials based only on a single observable, and it is necessary to compare and describe as many (related) observables as possible within the same model. To enable a comprehensive adjustment of models and new approaches which are under development \cite{Petersen:2018jag,Steinberg:2018jvv,Gallmeister:2017ths,Aichelin:2018jwh} to our data, differential transverse mass versus rapidity plots are shown in the Appendix.\\
\begin{figure}[tb]
	\begin{center}
		\resizebox{8.7cm}{!}{%				
			\includegraphics{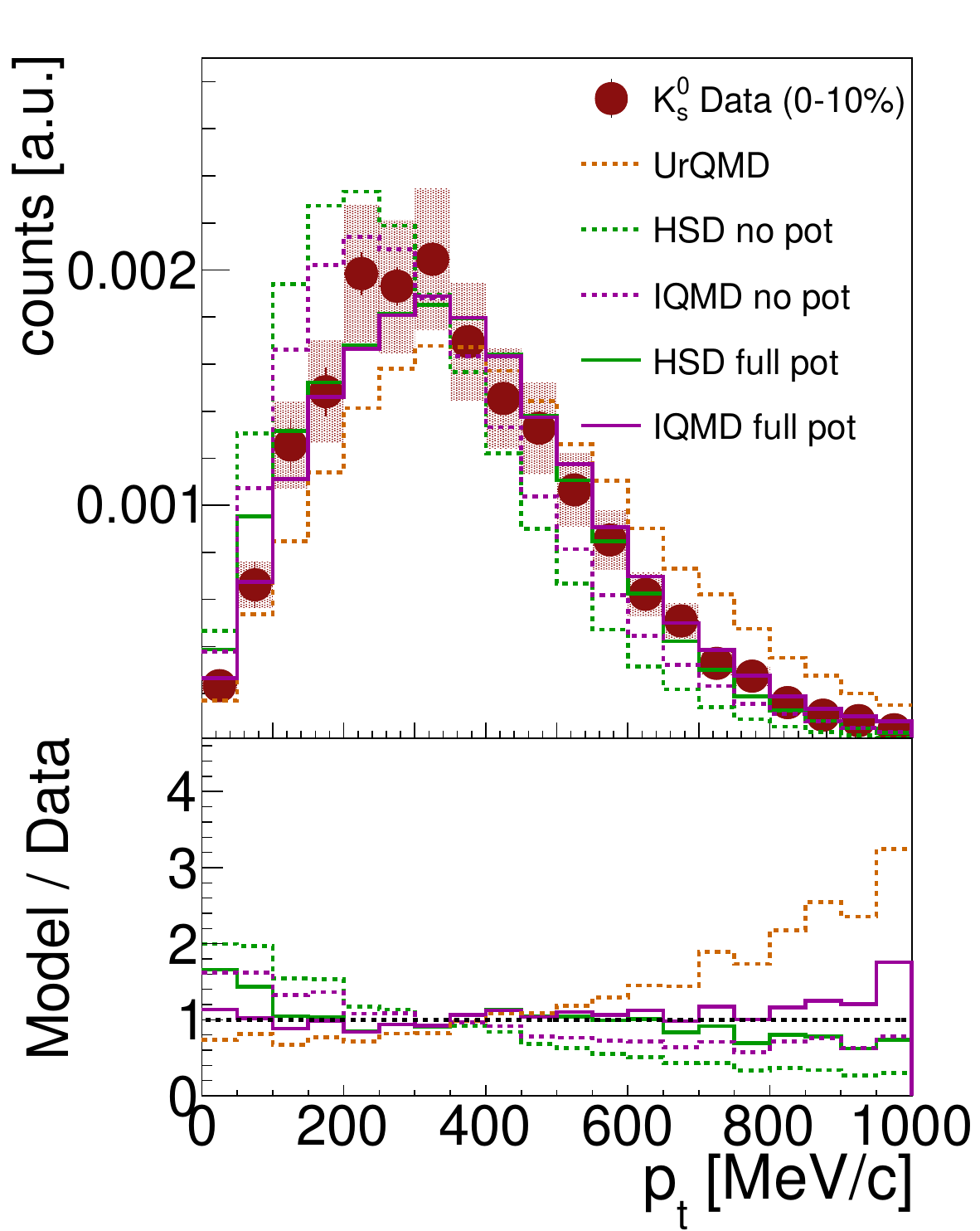}			
			\includegraphics{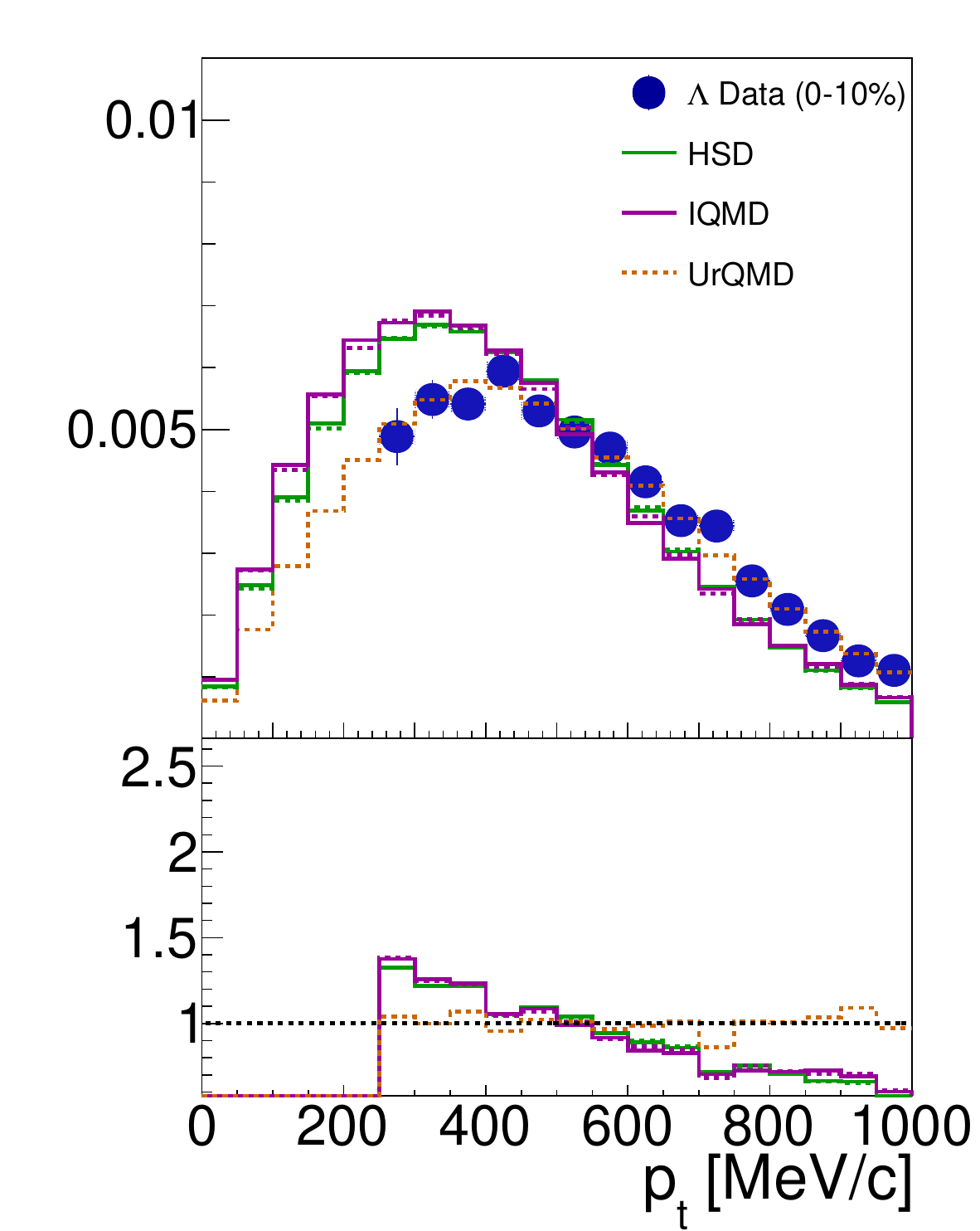}
			% epja impact_parameter_herb.eps!!
		}
	\end{center}
	\vspace{-0.7cm}  
	\caption{Comparison of the shape of the pt-spectra for $\pm$0.15 rapidity units around mid-rapidity of $K^0_s$ (left) and  $\Lambda$ (right) to various transport model versions. The model curves are normalized to the integral of the data.}
	\label{fig_pt}       % Give a unique label
\end{figure} 
%\section{Summary and Conclusion}
\begin{table}
	\begin{center}
	\begin{tabular}{|c|c|c|c|c|c|c|c|c|} \hline  
		\multirow{2}{*}{Model} & \multirow{2}{*}{KN potential} & \multicolumn{3}{c|}{$K^{0}_{s}$} & \multicolumn{3}{c|}{$\Lambda$} & \multirow{2}{*}{$\alpha$} \\\cline{3-8}
		& & $p_{t}$ & $y$ & Mult &  $p_{t}$ & $y$ & Mult & \\\hline      
		UrQMD & no & 105 & 4.1 & 1619 & 2.3 & 3.6 & 3020 & 16 \\\hline  
		HSD & yes & 7.0 & 2.7 & 670 & 39 &  6.3  & 626 & 6.3 \\\hline
		IQMD & yes & 6.0 & 2.0 & 99 & 38 & 12 & 214 & 0.3 \\\hline
	\end{tabular}
\end{center}
	\vspace{-0.5cm}
	\caption{Summary of the comparison of data to microscopic transport models based on the $\chi^2$ normalized to the number of data points.}
	\label{tabsum}
\end{table}
In summary, we present the first data on sub-threshold production of K$^0_s$ mesons and $\Lambda$ hyperons in Au+Au collisions at $\sqrt{s_{\rm{NN}}}=2.4$ GeV. We observe a universal scaling with $\langle A_{part} \rangle$ for all particles containing strangeness, independent of the corresponding excess energy. This suggests a more interrelated system than assumed in the past in which the total amount of strangeness increases stronger than linear with the number of participants and might be redistributed to the final hadron states only at freeze-out. Previous constraints on the EOS of nuclear matter based on the assumption of energy accumulation in sequential nucleon-nucleon collisions should therefore be revisited.\\
Our comparison of the yields, the universal scaling and the shapes of rapidity and $p_t$ spectra  to three microscopic transport models does not yet lead to a consistent picture. Including a repulsive KN potential the IQMD predictions are by far closest to the data, with a remaining deviation of the yields of the order of 10\% and an agreement within errors of the extracted values of $\alpha$. The shape of the kaon rapidity distribution is well described if a repulsive K-N potential is included in HSD and in particular in IQMD. On the other hand, UrQMD reproduces the rapidity distribution without such a potential, probably because of the particle production through intermediate resonances, but fails to reproduce the observed scaling of the yields with centrality. Yet, the shape of the $\Lambda$ rapidity distributions and the $p_t$ spectra are best described by UrQMD.
Due to the observed ambiguities and the imperfect description of the presented observables, it is premature to adjust the strength of the potential to a single observable. Further model refinements and subsequent data-to-model comparisons are necessary before firm constraints can be deduced.\\
{\bf Acknowledgment}\newline
The HADES collaboration thanks J. Aichelin, M. Bleicher, E. Bratkovskaya, C. Hartnack and J. Steinheimer for elucidating discussions. We gratefully acknowledge the support by the grants SIP JUC Cracow, Cracow (Poland), 2017/26/M/ST2/00600; TU Darmstadt, Darmstadt (Germany) and Goethe-University, Frankfurt (Germany), ExtreMe Matter Institute EMMI at GSI Darmstadt; TU M\"unchen, Garching (Germany), MLL M\"unchen, DFG EClust 153, GSI TMLRG1316F, BmBF 05P15WOFCA, SFB 1258, DFG FAB898/2-2; NRNU MEPhI Moscow, Moscow (Russia), in framework of Russian Academic Excellence Project 02.a03.21.0005, Ministry of Science and Education of the Russian Federation 3.3380.2017/4.6; JLU Giessen, Giessen (Germany), BMBF:05P12RGGHM; IPN Orsay, Orsay Cedex (France), CNRS/IN2P3; NPI CAS, Rez, Rez (Czech Republic), MSMT LM2015049, OP VVV CZ.02.1.01/0.0/0.0/16 013/0001677, LTT17003. 

%\FloatBarrier 

\FloatBarrier
%\newpage

\section{Appendix}
\begin{figure}
	% Use the relevant command for your figure-insertion program
	% to insert the figure file.
	% For example, with the option graphics use
	\begin{center}
		\resizebox{8.5cm}{!}{%
			\includegraphics{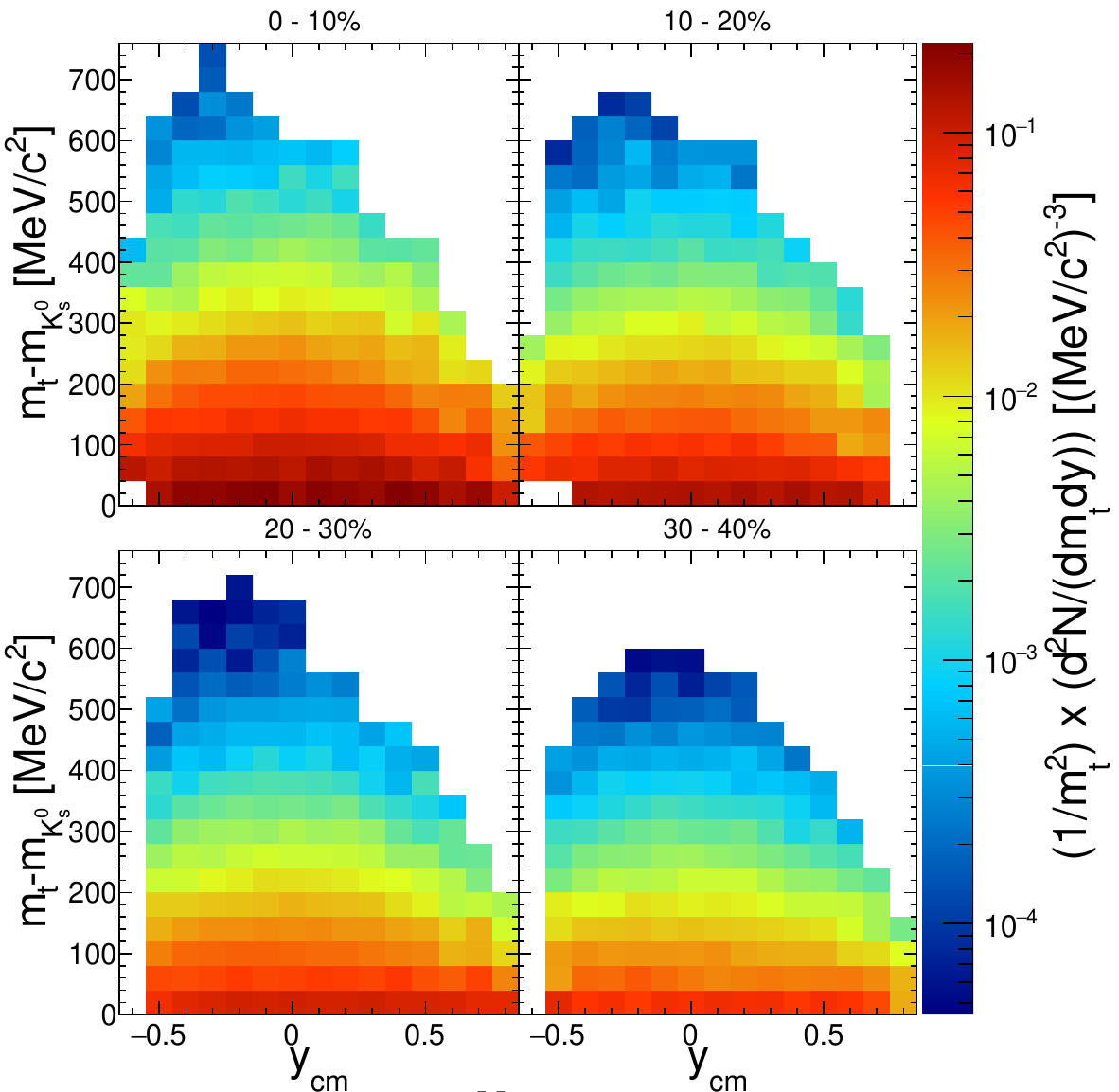}						
			% epja impact_parameter_herb.eps!!
		}
	\end{center}
	% If not, use
	\vspace{-0.7cm}       % Give the correct figure height in cm
	\caption{Differential yield of K$^0_s$ as function of rapdidity and reduced transverse mass for the four centrality classes.}
	\label{mtyK}       % Give a unique label  
\end{figure}

\begin{figure}
	% Use the relevant command for your figure-insertion program
	% to insert the figure file.
	% For example, with the option graphics use
	\begin{center}
		\resizebox{8.5cm}{!}{%
			\includegraphics{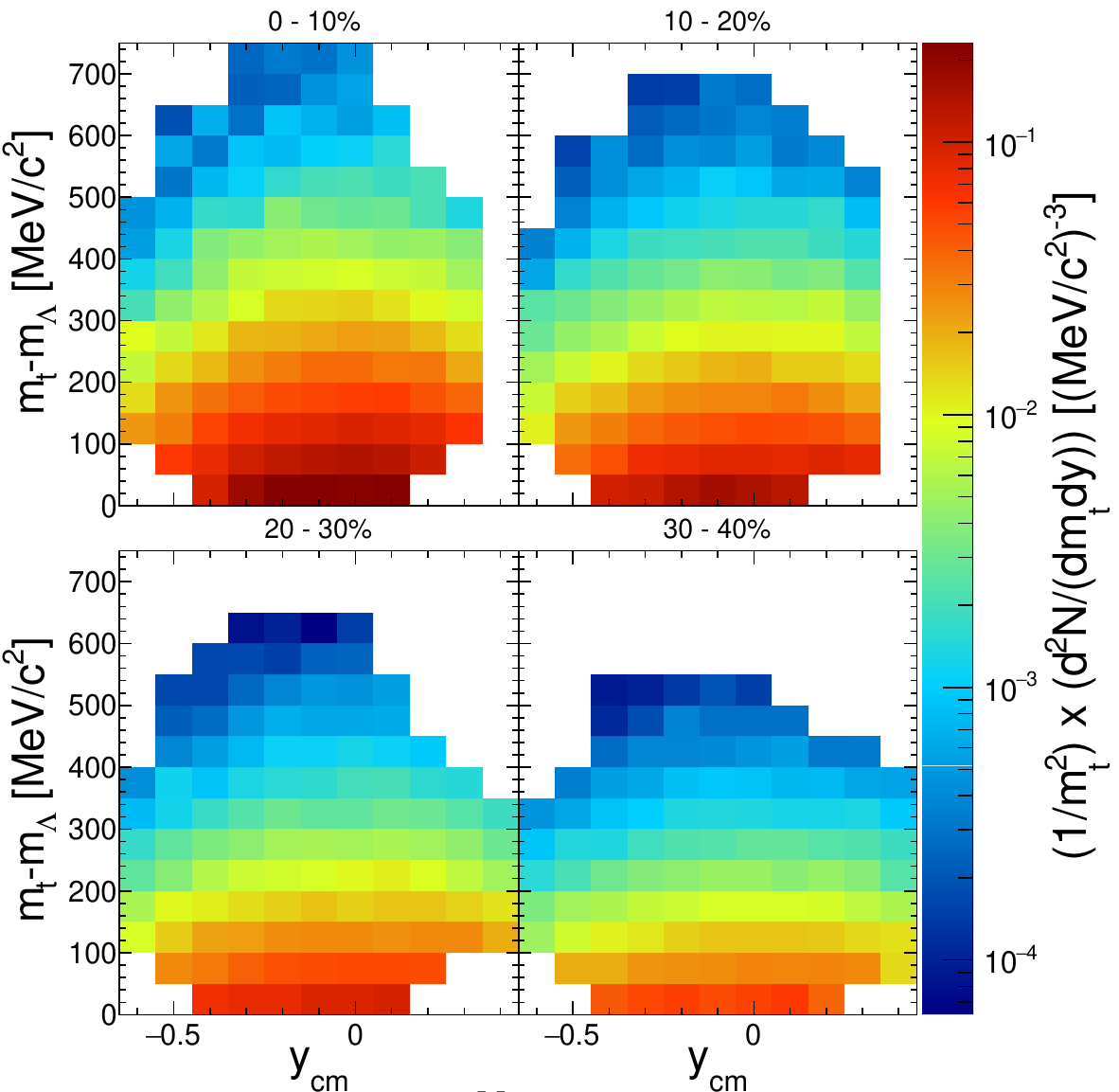}						
			% epja impact_parameter_herb.eps!!
		}
	\end{center}
	% If not, use
	\vspace{-0.7cm}       % Give the correct figure height in cm
	\caption{Differential yield of $\Lambda$  as function of rapidity and reduced transverse mass for the four centrality classes.}
	\label{mtyL}       % Give a unique label  
\end{figure}

The observables shown in the text are based on the data presented in Fig. \ref{mtyK} and \ref{mtyL}. The data are organized in bins of $m_{t}-m_{0}$ vs. $y_{cm}$ for four centrality classes. The details of the centrality selection are described in \cite{Kardan18}.
%\FloatBarrier

\end{document}